\documentclass[
 amsmath,amssymb,
prb,
twocolumn,
]{revtex4-2}
\usepackage{dcolumn}
\usepackage{bm}
\usepackage{braket}
\usepackage{color}
\usepackage{accents}
\usepackage{mathrsfs}
\usepackage{bm}
\usepackage{here}
\usepackage{longtable}

\usepackage{mathtools}
\usepackage{empheq}
\usepackage{graphicx}
\usepackage{hyperref}
\usepackage{comment}

\begin{document}
\title{
Finite-momentum superconducting states due to odd-frequency Cooper pairing correlations
}
\author{Takumi Sato$^{1}$}
\author{Satoru Hayami$^{1}$}
\author{Shingo Kobayashi$^{2}$}
\author{Yasuhiro Asano$^{3}$}%
\affiliation{
$^{1}$Graduate School of Science, Hokkaido University, Sapporo 060-0810, Japan\\
$^{2}$RIKEN Center for Emergent Matter Science, Wako, Saitama 351-0198, Japan\\
$^{3}$Department of Applied Physics, Hokkaido University, Sapporo 060-8628, Japan.\\
}

\date{\today}

\begin{abstract}
This paper discusses the origin of a nonuniform superconducting state in which Cooper pairs have a small but finite center-of-mass momentum. 
We analyze the instability of the normal state to such finite-momentum states using the pole of the pair fluctuation propagator in weak-coupling superconductors. 
The finite-momentum superconducting state is realized when the odd-frequency pairing correlations in the uniform superconducting state are expected to have sufficiently large amplitudes. 
We provide a perspective for a comprehensive understanding of inhomogeneous superconductivity and related phenomena.
\end{abstract}
\maketitle

\section{Introduction}
The Fulde-Ferrell-Larkin-Ovchinnikov (FFLO) state is a possible superconducting state 
in a conventional superconductor (SC) under a Zeeman field~\cite{fulde_fflo,larkin_fflo}.
In this state, Cooper pairs have a finite center-of-mass momentum $\bm{q}$ because a Zeeman field lifts the 
degeneracy of the Kramers partners.
As a result, the pair potential oscillates in real space with a period much longer than the 
Fermi wavelength.
In what follows, we refer to such a nonuniform superconducting state as a finite-momentum superconducting state. 
Time-reversal symmetry (TRS)-breaking fields in the normal state 
have been regarded as an important element for 
the finite-momentum superconducting states~\cite{sumita:prr2023,zhang:natcomm2024,chakraborty:prb2024}.
Indeed, the superconducting diode effect has 
been actively discussed as a related phenomenon to the FFLO state%
~\cite{yuan:pnas2022diode,daido:prl2022diode,he:njp2022diode,hasan:prb2024,shaffer:prb2024}.
However, other theoretical studies have suggested the possibility of 
the finite-momentum states in TRS-preserving SCs such as
noncentrosymmetric SCs and $j=3/2$ SCs~\cite{mineev:prb2008,li:prb2024}.
A common feature of these materials is that an electron has internal degrees of 
freedom such as spin, sublattice, band, and orbital.
Unfortunately, our understanding of this issue is still limited. 
For instance, the results for a $j=3/2$ SC suggest 
a positive correlation between the amplitude of interband Cooper pairs, 
which are composed of two electrons in different bands, and the emergence of finite-momentum states.
In contrast, another theory does not indicate such correlations~\cite{TS:prb2024}.
Thus, the role of such interband/interorbital Cooper pairs in forming finite-momentum states remains unclear.
A more comprehensive physical picture
that explains the mechanisms for stabilizing finite-momentum superconducting states is therefore desirable.
We address this issue in the present paper.

Another prominent example of finite-momentum superconductivity is the pair density wave (PDW) state proposed,
for example, in cuprates~\cite{agterberg:annrev2020,chakraborty:njp2021}.
Fermi surface nesting plays a crucial role in stabilizing such spatially oscillating 
superconducting states. 
As a result, the center-of-mass momentum of a Cooper pair is comparable to the Fermi wavenumber, 
(i.e., $q \sim k_{\mathrm{F}}$).
The PDW states are therefore beyond the scope of this paper because the mechanisms underlying them 
are qualitatively different from those discussed here.
In fact, we will show that the amplitude of the center-of-mass momentum in
finite-momentum superconducting states is much smaller than the Fermi 
wavenumber, (i.e., $q \ll k_{\mathrm{F}}$).

To clarify the problem, we summarize the common features among 
inhomogeneous superconducting states near a vortex core~\cite{tanuma:prl2009},
those near a magnetic impurity (cluster)~\cite{kuzmanovski:prb2020,perrin:prl2020,shu:prb2022,shu:jpscp2023,shu:prb2023},
and those at the surface of an unconventional SC~\cite{tanaka:prl2007,asano:prb2013}.
These theoretical studies have indicated the existence of 
odd-frequency Cooper pairs~\cite{berezinskii:jetplett1974,bergeret:rmp2005,
tanaka:jpsj2012,linder:rmp2019,triola:annphys2020,cayao:epj2020} 
around the inhomogeneous regions.
Odd-frequency Cooper pairs appear as induced subdominant pairing correlations by the local defects. 
In the bulk region, where the superconducting state is almost uniform, conventional even-frequency 
Cooper pairs form the pair potential and stabilize the superconducting state.
Namely, the pair potential itself belongs to conventional even-parity spin-singlet or 
odd-parity spin-triplet classes. Odd-frequency pairs do not form any pair
potentials because the corresponding attractive interactions are absent.
The odd-frequency pairing correlation functions always constitute part of the solution of the
Gor'kov equation in inhomogeneous SCs~\cite{asano:prb2014} and 
make the local superfluid density $n(\bm{r})$ negative
due to their paramagnetic property~\cite{asano:prl2011,suzuki:prb2014,higashitani:prb2014}.
The spatial variation of the pair potential increases the energy of
$\Delta E (\bm{r}) \approx n(\bm{r}) \, \bm{q}^2/m$ locally, where $\bm{q}$ and
$m$ are the momentum and the effective mass of a Cooper pair, respectively.
When the odd-frequency correlations locally reduce the superfluid density $n(\bm{r})$, 
the energy cost decreases by such amount.
Therefore, odd-frequency Cooper pairs represent and simultaneously support 
the local deformation of the superconducting condensate around a defect.
In what follows, we will show that this is also true for finite-momentum superconducting states in the bulk.
The existence of odd-frequency pairing correlations in the bulk was first pointed 
out in uniform multiband/orbital SCs~\cite{BSchaffer:prb2013}.
Odd-frequency correlations increase the free-energy and then decrease the
transition temperature of uniform superconducting states 
because of their paramagnetic response~\cite{asano:prb2015}.
The conclusions of these studies suggest that odd-frequency pairing correlations 
play a role in promoting the transition from a homogeneous superconducting state to an inhomogeneous one.

The mechanisms and the properties of finite-momentum superconducting states have already been
explored from various viewpoints%
~\cite{fulde_fflo,larkin_fflo,mineev:prb2008,li:prb2024,samokhin:prb2017,kitamura:prb2022FFLO}.
The aim of this paper is to provide another physical picture  
that explains the mechanisms underlying the emergence of finite-momentum superconducting 
states in various SCs.
To this end, we examine the instability of the normal state 
in the presence of an attractive interaction between two electrons~\cite{cooper:pr1956} 
by calculating the pair fluctuation propagator $D_{\mathrm{s}} (\bm{q})$ 
within the ladder approximation~\cite{ambegaokar:book1969}.
It is possible to calculate two transition temperatures from the poles of $D_{\mathrm{s}} (\bm{q})$: 
the transition temperature to a uniform superconducting state $T_{\bm{0}}$ and 
the transition temperature to a finite-momentum superconducting state $T_{\bm{q}}$. 
The superconducting state with a higher transition temperature is more stable.
To analyze the mechanisms stabilizing finite-momentum superconducting states with small $\bm{q}$, 
we expand the denominator of the fluctuation propagator as $D_{\mathrm{s}}^{-1} (\bm{q}) \propto a + B\, q^2=0$.
For $B>0$, we find that $T_{\bm{0}}>T_{\bm{q}}$, indicating the appearance of a uniform superconducting state.
For $B<0$, we find that $T_{\bm{0}}<T_{\bm{q}}$, indicating the appearance of 
a finite-momentum superconducting state with $\bm{q}$.
A general relationship shows that 
the coefficient $B$ is proportional to the Meissner kernel, or 
equivalently to the superfluid density $Q$, in a \textsl{uniform} superconducting state.
We conclude that the sign change of $Q$ due to the odd-frequency pairing correlations in a 
uniform state explains the instability of the uniform superconducting state and 
the emergence of finite-momentum states.

\subsection{Outline of this paper}
To clarify the logical flow leading to the conclusion, 
we explain in detail the role of each section of this paper. 

In Sec.~\ref{sec:odd}, we show the existence of odd-frequency pairing correlations in several SCs such as
a single-band SC with spin-dependent potentials, a two-band/two-orbital SC 
in the presence of band/orbital hybridization, and SCs in a vector potential.
We discuss the necessary conditions for the emergence of odd-frequency pairs.

In Sec.~\ref{sec:propagator},
we first derive a general expression for the pair fluctuation propagator $D_{\mathrm{s}} (\bm{q})$, 
which is a theoretical tool for analyzing the stability of finite-momentum superconducting states.
The transition temperature is calculated from the solutions of $D_{\mathrm{s}}^{-1} (\bm{q})=0$.
We expand the propagator up to the second order in $\bm{q}$ as $D_{\mathrm{s}}^{-1} (\bm{q})=a +B\, q^2$. 
The uniform superconducting state appears for $a=0$ and $B>0$, whereas  
the finite-momentum state with small $q$ is more stable for $a>0$ and $B<0$. 
Our analytic calculation shows a proportional relationship between the coefficient $B$ and the superfluid weight $Q$. 
Therefore, the finite-momentum superconducting states appear when 
the amplitude of the odd-frequency pairing correlations 
is large enough to change the sign of $Q$ in the corresponding uniform state.
This is the central conclusion of this paper.
Although the validity of the theory is limited to sufficiently small $q \ll k_{\mathrm{F}}$,
the condition for the emergence of finite-momentum superconducting states can be applied to any SC.

In Sec.~\ref{sec:FFLO},
we apply our theory to the well-known FFLO state in a conventional SC in Zeeman fields.
We demonstrate that the odd-frequency spin-triplet $s$-wave correlation decreases 
the superfluid weight by using the analytic expression of $Q$.
To check the validity of our theory in Sec.~\ref{sec:propagator}, we solve the equation 
$D_{\mathrm{s}}^{-1} (\bm{q})=0$ fully numerically and obtain the highest transition temperature 
$T_{\bm{q}}$ and the corresponding $\bm{q}$. 
The numerical results show a transition to the FFLO state in Zeeman fields large enough to make $Q<0$. 
All the numerical solutions for $\bm{q}$ in the FFLO state satisfy $q \ll k_{\mathrm{F}}$. 
These results support the main conclusion. 

In Sec.~\ref{sec:FFLO_multiband}, we apply our theory to two TRS-preserving SCs: 
a two-band SC with $s$-wave interband pairing order~\cite{silva:physletta2014} 
and a $j=3/2$ SC with $s$-wave pseudospin-quintet pairing order~\cite{brydon:prl2016}. 
By solving $D_{\mathrm{s}}^{-1} (\bm{q})=0$ numerically, we confirm that finite-momentum 
superconducting states with small $q$ become more stable than the uniform state.

In Sec.~\ref{sec:discussion}, we discuss the possibility of finite-momentum superconducting states in real materials. 
Moreover, we briefly discuss the close relationship between
the second-order transition to finite-momentum superconducting states 
and the first-order transition to the uniform state.
The conclusion is given in Sec.~\ref{sec:conclusion}.

Throughout this paper, we use the units of $k_\mathrm{B} = c = \hbar = 1$
where $k_\mathrm{B}$ is the Boltzmann constant and $c$ is the speed of light
and $e<0$ is the charge of an electron.

\section{Odd-frequency pairing correlations}
\label{sec:odd}
In this section, we show the general condition for the appearance of odd-frequency 
pairing correlations and apply the condition to single-band SCs with spin-dependent potentials
and to multiband/orbital SCs with band/orbital hybridization.   
The TRS-breaking fields have been considered as essential ingredients for 
realizing finite-momentum superconducting states.
In fact, extensive theoretical and experimental research has been devoted to 
exploring superconducting states that survive under strong magnetic fields
near the Pauli limit~\cite{chandrasekhar:apl1962,clogston:prl1962,matsuda:jpsj2007}.
Here, we reconsider the importance of TRS-breaking fields from the perspective of odd-frequency Cooper pairing. 

Generally speaking, the Bogoliubov-de Gennes (BdG) Hamiltonian for uniform superconducting states takes the form
\begin{align}
    H_{\mathrm{BdG}}(\bm{k}) = \left[
    \begin{array}{cc}
        \hat{\xi}_{\bm{k}} &  \hat{\Delta}(\bm{k}) \\
        -\hat{\Delta}^{\ast}(-\bm{k}) & -\hat{\xi}_{-\bm{k}}^\ast 
    \end{array} \right].
\end{align}
The anomalous Green's function, obtained as a solution of the Gor'kov equation, is formally expressed as
\begin{align}
    \hat{\mathcal{F}}(\bm{k},i\omega_{\ell}) &= \left[
        \hat{\Delta}(\bm{k}) \hat{\Delta}^{\ast}(-\bm{k}) - \omega_{\ell}^2 
        - \hat{\Delta}(\bm{k}) \hat{\xi}^{\ast}_{-\bm{k}} \hat{\Delta}^{-1}(\bm{k}) \hat{\xi}_{\bm{k}}
        \right. \nonumber \\
        &\hspace{1em}\left.-i\omega_{\ell} \hat{P}_{\mathrm{O}} (\bm{k}) \hat{\Delta}^{-1} (\bm{k})
    \right]^{-1} \hat{\Delta}(\bm{k}) , \\
    \hat{P}_{\mathrm{O}} (\bm{k}) &\coloneqq
    \hat{\xi}_{\bm{k}} \hat{\Delta}(\bm{k}) - \hat{\Delta}(\bm{k}) \hat{\xi}^{\ast}_{-\bm{k}}, 
    \label{eq:po_def}
\end{align}
where $\hat{P}_{\mathrm{O}}(\bm{k})$ characterizes the presence of odd-frequency pairing correlations%
~\cite{triola:annphys2020,ramires:prb2016,ramires:prb2018}.
When an attractive interaction works between the Kramers partners,
the pair potential in the BdG Hamiltonian can be expressed as
$\hat{\Delta}(\bm{k}) = \Delta(\bm{k})\,  \hat{U}_{T}$ 
where $\Delta(\bm{k})$ is an even function of $\bm{k}$ representing an even-parity superconductivity. 
The matrix $\hat{U}_T$ is the unitary part of the time-reversal 
operator $\mathcal{T}= \hat{U}_T \, \mathcal{C}$, 
where $\mathcal{C}$ denotes complex conjugation combined with the transformation $\bm{k} \to - \bm{k}$.
The relation
\begin{align}
    \hat{P}_{\mathrm{O}}(\bm{k}) 
    &= \left[
        \hat{\xi}_{\bm{k}} - \hat{U}_T \hat{\xi}^{\ast}_{-\bm{k}} \hat{U}_T^{-1}
    \right] \hat{\Delta}(\bm{k}) \nonumber \\
	&= \left[
        \hat{\xi}_{\bm{k}} - \mathcal{T} \,  \hat{\xi}_{\bm{k}}\,  \mathcal{T}^{-1}
    \right] \hat{\Delta}(\bm{k}) ,
    \label{eq:PO_Kramers}
\end{align}
indicates that TRS-breaking fields in the normal state Hamiltonian 
is necessary for the emergence of the odd-frequency pairing correlation.

To discuss the importance of TRS-breaking fields, we begin with spin-singlet 
even-parity SCs for spin $s=1/2$ electrons, where $\hat{U}_T$ is replaced by $i \hat{\sigma}_2$
with $\hat{\bm{\sigma}}=(\hat{\sigma}_1, \hat{\sigma}_2, \hat{\sigma}_3)$ being 
the Pauli matrices in spin space.
The normal state Hamiltonian includes two types of spin active potentials:
\begin{align}
\hat{\xi}_{\bm{k}} &= \xi_{\bm{k}} + \bm{h}_{\bm{k}} \cdot \hat{\bm{\sigma}} + \bm{\lambda}_{\bm{k}} \cdot \hat{\bm{\sigma}}. \label{eq:hn_s12}
\end{align}
The TRS-breaking fields $\bm{h}_{\bm{k}} = \bm{h}_{-\bm{k}}$ represent a Zeeman field, 
a ferromagnetic exchange field%
~\cite{chandrasekhar:apl1962,clogston:prl1962,sarma:jpcs1963,maki_tsuneto:ptp1964,bergeret:prl2001}, and  
an altermagnetic exchange field%
~\cite{noda:pccp2016,okugawa:jpcm2018,smejkal:sciadv2020,naka:natcomm2019,ahn:prb2019, 
hayami:jpsj2019spinsplit,hayami:jpscp2020,hayami:prb2020spinsplit,yuan:prb2020,naka:prb2021perovskite, 
yuan:prm2021,yuan:prb2021,hernandez:prl2021,smejkal:prx2022magnetoresistance,mazin:pnas2021, 
smejkal:prx2022spingroup1,smejkal:prx2022spingroup2, 
cheong:npj2025AMclassification,guo:advmat2025review,hu:prx2025}.
On the contrary, the potential $\bm{\lambda}_{\bm{k}}= - \bm{\lambda}_{-\bm{k}}$ 
represents antisymmetric spin-orbit 
interactions preserving TRS~\cite{gorkov:prl2001,frigeri:prl2004,frigeri:njp2004}.
The relation in Eq.~\eqref{eq:PO_Kramers} shows that only $\bm{h}_{\bm{k}}$ induces the odd-frequency
Cooper pairs in the spin-singlet SCs, as listed in the second row of Table~\ref{table1}. 

The TRS-breaking fields become less important in spin-triplet SCs described by  
the pair potential
\begin{align}
    \hat{\Delta}(\bm{k}) &=  \bm{d}_{\bm{k}} \cdot \hat{\bm{\sigma}} \,  i\, \hat{\sigma}_2 , 
\end{align}
with $\bm{d}_{\bm{k}} = -\bm{d}_{-\bm{k}}$.
In a spin-triplet SC, two types of odd-frequency correlations exist, as listed in the 
third row of Table~\ref{table1}.
One is the spin-singlet odd-parity correlation induced by the TRS-breaking field
$\bm{h}_{\bm{k}}$
and the other is the spin-triplet even-parity correlation induced by 
the spin-orbit interactions $\bm{\lambda}_{\bm{k}}$ preserving TRS. 
The authors in Refs.~\cite{frigeri:prl2004,frigeri:njp2004} show that
 $\bm{\lambda}_{\bm{k}} \times \bm{d}_{\bm{k}}=0$ 
is necessary for realizing a stable superconducting state.
Namely, $T_c$ for $\bm{\lambda}_{\bm{k}} \times \bm{d}_{\bm{k}} =0$ is higher than 
that for $\bm{\lambda}_{\bm{k}} \times \bm{d}_{\bm{k}} \neq 0$.
The instability (stability) of superconducting states is explained well 
by the presence (absence) of odd-frequency Cooper pairs in the bulk~\cite{asano:prb2015}.

The TRS-breaking fields are not a necessary item for the odd-frequency pairing correlations
when an electron has extra internal degrees of freedom. 
To clarify this point, we consider a two-band SC, 
where the Pauli matrices $\hat{\rho}_{\mathrm{i}}$ for $\mathrm{i}=1-3$ describe 
the $2\times 2$ band space in the normal state Hamiltonian 
\begin{align}
\hat{\xi}_{\bm{k}} &= \xi_{\bm{k}} \hat{\rho}_0 +  V \hat{\rho}_1 + V^\prime \hat{\rho}_2 + 
\varepsilon \hat{\rho}_3.
\label{eq:hn_tb}
\end{align}
The potentials $V$ and $V'$ represent the band hybridization, and $\varepsilon$ represents band asymmetry.
The spin degree of freedom is omitted because the normal state Hamiltonian contains no spin active terms.
The interband pair potential described by $\Delta\, \hat{\rho}_1$ represents
a spin-singlet $s$-wave even-band-parity pair potential.
In this case, Eq.~\eqref{eq:po_def} becomes $\hat{P}_\mathrm{O} = 2 i \Delta \varepsilon \hat{\rho}_2$, 
which indicates the emergence of odd-frequency odd-band-parity pairing correlations~\cite{BSchaffer:prb2013}.
In Table.~\ref{table2}, we summarize the matrix structures of the pair potentials and 
those of the induced odd-frequency correlations. 
For all even-band-parity pair potentials such as $\Delta \hat{\rho}_0$, $\Delta \hat{\rho}_1$, 
and $\Delta \hat{\rho}_3$, the band hybridization or asymmetry 
generates odd-frequency correlations belonging to the odd-band-parity symmetry class. 
When we consider an odd-band-parity pair potential, $\Delta \hat{\rho}_2$, 
the induced odd-frequency correlations belong to the even-band-parity class~\cite{BSchaffer:prb2013}.
Therefore, the potentials that hybridize the internal 
degrees of freedom of electrons are crucial for the emergence of the odd-frequency pairing 
correlations.

Finally, we point out a trivial potential that generates odd-frequency pairing correlations.
In the presence of a vector potential, the normal state Hamiltonian is given by
\begin{align}
    \hat{\xi}_{\bm{k}} &=
    \xi_{\bm{k}} + \frac{e^2 \bm{A}^2}{2m} - \frac{e\hbar}{m} \bm{A} \cdot \bm{k} .
\end{align}
Since a vector potential couples to the charge of an electron, 
$\hat{\xi}_{\bm{k}}$ is always proportional to the identity matrix in any internal spaces. 
We find that odd-frequency correlations characterized by $\hat{P}_\mathrm{O} \propto e \bm{A} \cdot \bm{k}$ 
are induced, belonging to the parity opposite to that of the pair potential.

\begin{table}[t]
    \caption{
        The pair potentials and the induced odd-frequency pairing correlations are 
        summarized for the single band $s=1/2$ SCs.
        The normal state Hamiltonian is given in Eq.~\eqref{eq:hn_s12}.
    }
    \begin{ruledtabular}
        \renewcommand{\arraystretch}{1.3}
        \begin{tabular}{ll}
            Pair potential & $\hat{P}_\mathrm{O}$ \\
            \colrule
            $\Delta_{\bm{k}} \, i\hat{\sigma}_2$ &
            $2 \Delta_{\bm{k}} \bm{h}_{\bm{k}} \cdot \hat{\bm{\sigma}}  i\hat{\sigma}_2$ \\
            $\bm{d}_{\bm{k}} \cdot \hat{\bm{\sigma}} \, i\hat{\sigma}_2$ & 
            $2 [ \bm{d}_{\bm{k}} \cdot \bm{h}_{\bm{k}} + i (\bm{\lambda}_{\bm{k}} \times \bm{d}_{\bm{k}})
            \cdot \hat{\bm{\sigma}}   ] i\hat{\sigma}_2$ 
        \end{tabular}
    \end{ruledtabular}
    \label{table1}
\end{table}

\begin{table}[t]
    \caption{
        The pair potentials and the induced odd-frequency pairing correlations are 
        summarized for the two-band SCs.
        The normal state Hamiltonian is given in Eq.~\eqref{eq:hn_tb}.
    }
    \begin{ruledtabular}
        \renewcommand{\arraystretch}{1.3}
        \begin{tabular}{ll}
            Pair potential & $\hat{P}_\mathrm{O}$ \\
            \colrule
            $\Delta\, \hat{\rho}_0$ & $2 \Delta V^\prime \hat{\rho}_2$ \\
            $\Delta\, \hat{\rho}_1$ & $2 i \Delta \varepsilon \hat{\rho}_2$ \\
            $\Delta\, \hat{\rho}_2$ & $2 \Delta ( V^\prime \hat{\rho}_0- i \varepsilon \hat{\rho}_1
            + iV \hat{\rho}_3)$ \\
            $\Delta\, \hat{\rho}_3$ & $-2 i \Delta  V \hat{\rho}_2$
        \end{tabular}
    \end{ruledtabular}
    \label{table2}
\end{table}

\section{Pair fluctuation propagator}
\label{sec:propagator}
In this section, we introduce a theoretical tool to discuss the instability of the normal 
state in the presence of attractive interactions between two electrons.
The Hamiltonian describing such electronic states is given by
\begin{align}
    \mathcal{H} &= \mathcal{H}_0 + \mathcal{H}_1
    \label{eq:h} , \\
    \mathcal{H}_0 &= \sum_{\bm{k}\alpha\beta} \xi_{\bm{k}\alpha\beta} c^{\dag}_{\bm{k}\alpha} c_{\bm{k}\beta} , \quad
    \label{eq:h0k}
    \xi_{\bm{k}\alpha\beta} = \epsilon_{\bm{k}\alpha\beta} - \mu \delta_{\alpha\beta} , \\
    \mathcal{H}_1 &= -g \sum_{\bm{q}} \Phi^{\dag}_{\bm{q}} \Phi_{\bm{q}} , \\
    \Phi_{\bm{q}} &= \frac{1}{\sqrt{2N}} \sum_{\bm{k}'\gamma\delta} w^{\ast}_{\gamma\delta}(\bm{k}')
    c_{\bm{k}'+\frac{\bm{q}}{2}\gamma} c_{-\bm{k}'+\frac{\bm{q}}{2}\delta} ,
\end{align}
where $\alpha,\beta,\gamma,\delta$ denote the internal degrees of freedom of an electron,
such as spin, band, orbital, and sublattice, 
$c_{\bm{k}\alpha}$ is the annihilation operator of an electron with $\bm{k}$ and $\alpha$,  
and $g>0$ represents the strength of the attractive interaction.
$\Phi_{\bm{q}}$ is the annihilation operator of a Cooper pair with center-of-mass momentum $\bm{q}$, 
and $N$ is the number of unit cells in the underlying lattice.
The attractive interaction is characterized by $w_{\sigma\sigma'}(\bm{k})$, which satisfies
$w_{\sigma\sigma'}(\bm{k}) = -w_{\sigma'\sigma}(-\bm{k})$ due to the fermionic anticommutation relations.

Within the linear response theory~\cite{ambegaokar:book1969},
the transition from the normal state to the superconducting state is examined
by analyzing the pole of the pair fluctuation propagator, which is defined as
\begin{align}
    D_{\mathrm{s}} (\bm{q},i\nu_n) &\coloneqq
    -\int_{0}^{1/T} d\tau \, \langle T_{\tau} \Phi_{\bm{q}} (\tau) \Phi^{\dag}_{\bm{q}} \rangle \,  e^{i\nu_n \tau},
\end{align}
with $\Phi_{\bm{q}} (\tau) = e^{\tau\mathcal{H}} \Phi_{\bm{q}} e^{-\tau\mathcal{H}}$, 
where $\nu_n=2n\pi T$ is the bosonic Matsubara frequency with $n$ and $T$ being an integer number 
and a temperature, respectively.
By summing the ladder diagrams, we obtain
\begin{align}
    D_{\mathrm{s}} (\bm{q},i\nu_n) &= \frac{-\Pi_{\mathrm{s}0} (\bm{q},i\nu_n)}{1-g\Pi_{\mathrm{s}0} (\bm{q},i\nu_n)} , \label{eq:ds_def}\\
    \Pi_{\mathrm{s}0} (\bm{q},i\nu_n) &=
    -T\sum_{\omega_{\ell}} \frac{1}{N} \sum_{\bm{k}}
    \mathrm{Tr} \left[
    \hat{\mathcal{G}}_0 (\bm{k}+\bm{q}/2, i\omega_{\ell})\,  \hat{w}(\bm{k}) \right. \nonumber \\
     \times &\left.
    \hat{\underline{\mathcal{G}}}_0 (\bm{k}-\bm{q}/2, i\omega_{\ell}-i\nu_n) \, \hat{w}^{\dag}(\bm{k})
    \right] \label{eq:pi_def},
\end{align}
where $\hat{\mathcal{G}}_0(\bm{k},i\omega_{\ell})=[i\omega_{\ell}-\hat{\xi}_{\bm{k}}]^{-1}$
is the Green's function in the absence of the attractive interaction,
$\underline{X}(\bm{k},i\omega_{\ell}) \coloneqq -X^{\ast}(-\bm{k},i\omega_{\ell})$ represents
particle-hole conjugation of a function $X(\bm{k},i\omega_{\ell})$
and $\omega_{\ell}=(2\ell + 1)\pi T$ is the fermionic Matsubara frequency with $\ell$ being an integer number.
The second-order transition to a superconducting phase is characterized by the divergence of the retarded propagator 
$D^{\mathrm{R}}_{\mathrm{s}} (\bm{q},\omega) \coloneqq D_{\mathrm{s}} (\bm{q},i\nu_n \rightarrow \omega + i\delta)$.
In the following, we put $\omega= 0$, since we focus on static superconducting states.
To discuss the transition to finite-momentum superconducting states with small $q$, 
we expand the pair polarization function
$\Pi^{\mathrm{R}}_{\mathrm{s}0} (\bm{q},0) \coloneqq
\Pi_{\mathrm{s}0} (\bm{q},i\nu_n \rightarrow \omega + i\delta = 0)$ with respect to $\bm{q}$:
\begin{widetext}
\begin{align}
    \Pi^{\mathrm{R}}_{\mathrm{s}0} (\bm{q},0) &=
    -T\sum_{\omega_{\ell}} \frac{1}{N} \sum_{\bm{k}} \mathrm{Tr}
    \bigg[
        \hat{\mathcal{G}}_0 \hat{w} \hat{\mathcal{\underline{G}}}_0 \hat{\underline{w}} \nonumber \\
        &-\frac{q_{\mu}q_{\nu}}{4} \mathrm{Re} \left[
            \partial_{k_{\mu}} \partial_{k_{\nu}} \hat{\xi} \, \hat{\mathcal{G}}_0 \, \hat{w} 
			\hat{\mathcal{\underline{G}}}_0 \, \hat{\underline{w}} \, 
            \hat{\mathcal{G}}_0 
            +\hat{v}_{\mu} \, \hat{\mathcal{G}}_0\,  \hat{v}_{\nu} \, 
			\hat{\mathcal{G}}_0 \, \hat{w} \, \hat{\mathcal{\underline{G}}}_0 \, \hat{\underline{w}}
            \hat{\mathcal{G}}_0
            +\hat{v}_{\mu}\,  \hat{\mathcal{G}}_0 \, \hat{w} \, \hat{\mathcal{\underline{G}}}_0 \, 
			\hat{\underline{w}} \, \hat{\mathcal{G}}_0 \, \hat{v}_{\nu} \,
            \hat{\mathcal{G}}_0
            +\hat{v}_{\mu}\,  \hat{\mathcal{G}}_0 \, \hat{w} \, \hat{\mathcal{\underline{G}}}_0 \, 
			\hat{\underline{v}}_{\nu} \, \hat{\mathcal{\underline{G}}}_0 \,  \hat{\underline{w}} \,
            \hat{\mathcal{G}}_0 \right] \nonumber\\
        &+ O(q^4) \bigg]
    \label{eq:Pi} ,
\end{align}
\end{widetext}
where the repeated indices $\mu,\nu=x,y,z$ are summed over,
$\partial_{k_{\mu}} fg = \frac{\partial f}{\partial k_{\mu}} g$,
$\hat{v}_{\mu} \coloneqq \partial_{k_{\mu}} \hat{\xi}$ is the velocity operator, and
$\hat{\mathcal{G}}_0$, $\hat{w}$, $\hat{\underline{w}}$, and $\hat{\underline{v}}_{\mu}$ are abbreviations of
$\hat{\mathcal{G}}_0 (\bm{k},i\omega_{\ell})$, $\hat{w} (\bm{k})$,
$-\hat{w}^{\ast}(-\bm{k})$, and $-\hat{v}^{\ast}_{\mu}(-\bm{k})$, respectively.
We adopt the lattice constant as the length unit. 
In Eq.~\eqref{eq:Pi}, we have assumed that the odd-order terms with respect to $\bm{q}$ vanish for simplicity
\footnote{
    The odd-order terms would play an important role in discussing other exotic phenomena, 
    such as the superconducting diode effect~\cite{yuan:pnas2022diode,daido:prl2022diode,he:njp2022diode}. 
    The effects of these terms will be discussed elsewhere.
}.

After some algebra, the pair fluctuation propagator can be expressed 
in terms of the Meissner kernel in the uniform superconducting state:
\begin{align}
    \left[ D^{\mathrm{R}}_{\mathrm{s}}(\bm{q},0)\right]^{-1} &\propto
    \frac{1}{g}- \Pi^{\mathrm{R}}_{\mathrm{s}0} (\bm{q},0) \nonumber \\
    &=a(T)+B_{\mu\nu}(T) q_{\mu}q_{\nu} + O(q^4) 
    \label{eq:DsR}, \\
    a(T) &= \frac{1}{g} + T\sum_{\omega_{\ell}} \frac{1}{N} \sum_{\bm{k}} \mathrm{Tr}
    \left[ \hat{\mathcal{G}}_0 \hat{w} \hat{\mathcal{\underline{G}}}_0 \hat{\underline{w}} \right] , \\
    B_{\mu\nu}(T)
    &= \left. \frac{K_{\mu\nu}(\bm{0},0)}{4e^2 \alpha^2|\Delta|^2} \right|_{\Delta=0}
    \label{eq:B} ,
\end{align}
where $a(T)$ and $K_{\mu\nu}(\bm{0},0)$ correspond to the quadratic coefficient
of the Ginzburg-Landau (GL) free-energy and the Meissner kernel for a uniform superconducting state, respectively.
The last relation between $B_{\mu\nu}(T)$ and $K_{\mu\nu}(\bm{0},0)$ is particularly important to 
understand how odd-frequency pairing correlations destabilize a uniform superconducting state.
The derivation of the Meissner kernel is presented in Appendix.~\ref{sec:meissner}.
According to the definition in Eq.~\eqref{eq_a:defdelta},
the pair potential can be represented as 
\begin{align}
\hat{\Delta}(\bm{k}) = \alpha \, \Delta \, \hat{w}(\bm{k}),\label{eq:del5}
\end{align}
with constants $\alpha \in \mathbb{R}$ and $\Delta \in \mathbb{C}$.  
By substituting the expression in Eq.~\eqref{eq:del5} into Eq.~\eqref{eq:Kexpand} 
and comparing the result with the second term in Eq.~\eqref{eq:Pi},  
we obtain the relation in Eq.~\eqref{eq:B}. 

Moreover, without loss of generality, the GL free-energy of a 
uniform superconducting state can be written as
\begin{align}
    \Omega_{\mathrm{GL}}(\Delta) = a(T)\, |\Delta|^2 + b(T) \, |\Delta|^4 + O(\Delta^6)
    \label{eq:Omega_GL} .
\end{align}
The transition temperature to a uniform superconducting state $T_{\bm{0}}$ is given by
\begin{align}
    a(T_{\bm{0}}) = 0 ,
\end{align}
since $a(T)$ changes the sign at $T=T_{\bm{0}}$.
The kind of the transition is characterized by the sign of $b(T_{\bm{0}})$.
The transition to the uniform superconducting state is 
a second-order (continuous) for $b>0$, whereas 
it is a first-order (discontinuous) for $b<0$.

Using Eq.~\eqref{eq:DsR}, the transition temperature to 
a finite-momentum state $T_{\bm{q}}$ is calculated from
\begin{align}
a(T_{\bm{q}})+B_{\mu\nu}(T_{\bm{q}}) q_{\mu}\, q_{\nu}=0 . \label{eq:tqdef}
\end{align}
We assume that $q$ is much smaller than $k_{\mathrm{F}}$ so 
that fourth- and higher-order terms can be neglected. 
To satisfy Eq.~\eqref{eq:tqdef}, the two cases are possible:
\begin{itemize}
    \item[(i)]  $a(T_{\bm{q}})<0$ and $B_{\mu\nu}(T_{\bm{q}}) \, q_\mu \, q_\nu>0$ ,
    \item[(ii)] $a(T_{\bm{q}})>0$ and $B_{\mu\nu}(T_{\bm{q}}) \, q_\mu \, q_\nu<0$ .
\end{itemize}
Case (i) gives $T_{\bm{q}}< T_{\bm{0}}$, indicating that
the uniform superconducting state is more stable than the finite-momentum states.
As a result, the transition to the uniform superconducting state occurs at $T=T_{\bm{0}}$.
Case (ii) gives $T_{\bm{q}} > T_{\bm{0}}$, indicating that
the transition to a finite-momentum state occurs at $T=T_{\bm{q}}$.
In this way, the stability of the inhomogeneous superconducting state is determined.
These conclusions hold true 
as long as the higher-order terms $O(q^4)$ in Eq.~\eqref{eq:DsR} are negligible.
The even- (odd-) frequency pairing correlations contribute positively (negatively) 
to the Meissner kernel in Eq.~\eqref{eq:B}~\cite{asano:prb2015}. 
Therefore, the stability of a superconducting state is determined by the relative amplitude 
between the even- and the odd-frequency pairing correlations.
We also find that $b$ in Eq.~\eqref{eq:Omega_GL} and  
$B_{\mu \nu}$ have the same sign in many cases.
This suggests a close relationship between the continuous transition to a finite-momentum 
state and the discontinuous transition to a uniform state. 
We will discuss this issue in Sec.~\ref{sec:discussion}.

At the end of this section, we briefly explain how the theory in this section 
connects with the results in the following sections.
In Sec.~\ref{sec:FFLO}, we apply the argument to the FFLO state in a spin-singlet 
$s$-wave SC in Zeeman fields. 
The transition to the FFLO state due to the TRS-breaking field is described well by the theory.
Although we neglect higher order terms in Eq.~\eqref{eq:DsR}, fully numerical solution 
of $[D_{\mathrm{s}}^{\mathrm{R}}]^{-1}=0$ gives finite-momentum states with $q \ll k_{\mathrm{F}}$. 
In Sec.~\ref{sec:FFLO_multiband}, we seek the possibility of finite-momentum states 
in two-band SCs and $j=3/2$ SCs that preserve TRS. 
The numerical results indicate the presence of finite-momentum superconducting state with $q \ll k_{\mathrm{F}}$.

\section{FFLO state}
\label{sec:FFLO}

\begin{figure*}[tbp]
    \includegraphics[width=\linewidth]{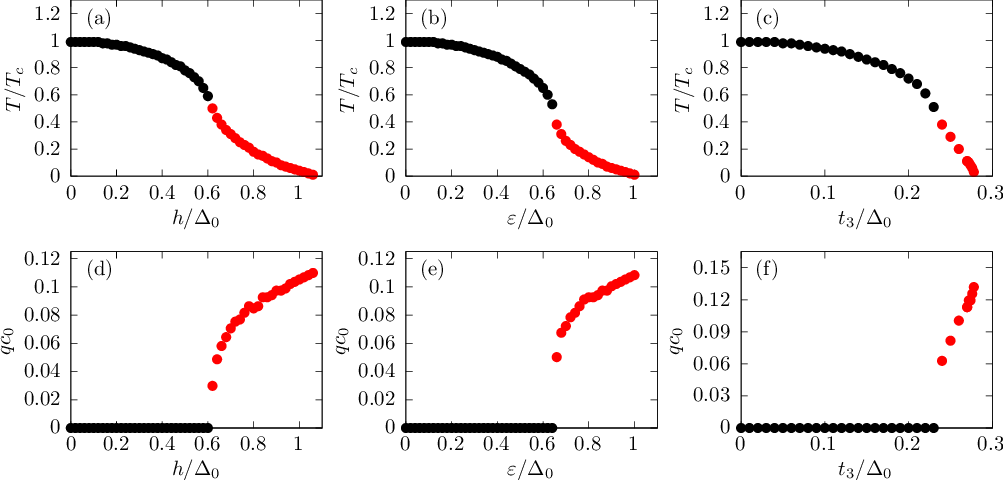}
    \caption{
        The phase boundary between the normal and superconducting states is shown in (a)--(c),
        which is determined by the zeros of the denominator of the pair fluctuation propagator 
		in Eq.~\eqref{eq:ds_def} with Eq.~\eqref{eq:pi_def}.
        Panels (a), (b), and (c) show the results for
        a conventional SC under Zeeman fields in the $h$--$T$ plane,
        a two-band SC with interband spin-singlet pairing order $(s=-1)$ in the $\varepsilon$--$T$ plane,
        and a $j=3/2$ SC in the $t_3$--$T$ plane, respectively.
        The transition points to the uniform (finite-momentum) state
        are indicated by black (red) dots.
        Figures (d)--(f) show the center-of-mass momentum of the Cooper pair $q$ at the transition point, 
        and the dots are colored in the same manner as in (a)--(c).
    }
    \label{fig:boundary_q}
\end{figure*}

In this section, we apply the theory in Sec.~\ref{sec:propagator}
to the FFLO state in a conventional SC~\cite{fulde_fflo,larkin_fflo}.
We consider spin-singlet superconductivity under Zeeman fields on a square lattice.
In Eq.~\eqref{eq:h}, we choose
\begin{align}
    \hat{\xi}_{\bm{k}} &= \xi_{\bm{k}} \hat{\sigma}_0 - \bm{h} \cdot \hat{\bm{\sigma}} , \quad
    \hat{w}(\bm{k}) = \frac{1}{\sqrt{2}} (i\hat{\sigma}_2) , \label{eq:hn_zeeman}\\
    \xi_{\bm{k}} &= -2t(\cos k_x + \cos k_y) + 4t - \mu ,
\end{align}
where $\bm{h} \coloneqq \mu_{\mathrm{B}} \bm{B}$ represents the Zeeman term,
and $\hat{\sigma}_0$ is the unit matrix.
The length is measured in units of the lattice constant $c_0=1$.
The pair potential in the BdG Hamiltonian for the uniform phase is given by
$\hat{\Delta}(\bm{k})=\Delta (i\hat{\sigma}_2)$.
To proceed with analytic calculations, we consider the continuum limit of the model:
$\xi_{\bm{k}} \rightarrow (k_x^2+k_y^2)/2m - \mu$ with $m = 1/2t$ and
$\frac{1}{N} \sum_{\bm{k}} \rightarrow \frac{1}{V_{\mathrm{vol}}} \sum_{\bm{k}}$, where $V_{\mathrm{vol}}$ denotes the volume of the system. 
By solving the Gor'kov equation in Eq.~\eqref{eq:gorkov}, the anomalous Green's function is obtained as~\cite{TS:prb2024}
\begin{align}
    &\hspace{1em}\hat{\mathcal{F}}_{\mathrm{z}}(\bm{k}, i\omega_{\ell}) = \nonumber \\
    &\hspace{2em}
    \frac{-1}{Z_{\mathrm{z}}}\left[ \omega_{\ell}^2 + 
    \xi_{\bm{k}}^2 + \Delta^2 - h^2
    +2 i \omega_{\ell} \, \bm{h} \cdot \hat{\bm{\sigma}} \right]
    \Delta ( i \hat{\sigma}_2 ) , \label{eq:F-func_zeeman} \\
    &\hspace{1em}Z_{\mathrm{z}} = \xi_{\bm{k}}^4 + 2\, \xi_{\bm{k}}^2\, A_{\mathrm{z}} + C_{\mathrm{z}}
    \label{eq:Zzeeman}, \\
    &\hspace{1em}A_{\mathrm{z}} =
    \omega_{\ell}^2 - h^2 +\Delta^2 , \quad
    C_{\mathrm{z}} = A^2_{\mathrm{z}} + 4 \omega_{\ell}^2\, h^2 ,
\end{align}
where the last term in Eq.~\eqref{eq:F-func_zeeman} represents the odd-frequency pairing correlation
induced by the Zeeman field~\cite{bergeret:prl2001,chakraborty:prb2022}.
The expression for the normal Green's function $\hat{\mathcal{G}}_{\mathrm{z}}$ is presented in Appendix~\ref{sec:g-func}.
Using the obtained Green's function,
the GL coefficient $a_{\mathrm{z}}(T)$ and the Meissner kernel $K_{{\mathrm{z}},\mu\nu}$ are calculated as~\cite{TS:prb2024}
\begin{align}
    &a_{\mathrm{z}}(T) = \frac{1}{g} - N_0 \pi T \sum_{\omega_{\ell}} 
    \frac{\omega_{\ell}^2}{|\omega_{\ell}| (\omega_{\ell}^2 + h^2)} , \label{eq:GL_zeeman} \\
    &K_{{\mathrm{z}},\mu\nu} (\bm{0},0) = \delta_{\mu\nu} \frac{2ne^2}{m} Q_{\mathrm{z}} , \label{eq:K_zeeman} \\
    &Q_{\mathrm{z}} = \sqrt{2} \pi T \sum_{\omega_{\ell}}
    \frac{\Delta^2 \left\{ A_{\mathrm{z}}^3+\sqrt{C_{\mathrm{z}}}(A_{\mathrm{z}}^2-2\omega_{\ell}^2 h^2) \right\} }
    {[ C_{\mathrm{z}} (A_{\mathrm{z}}+\sqrt{C_{\mathrm{z}}})]^{3/2}} . \label{eq:Q_zeeman}
\end{align}
where $n=k_{\mathrm{F}}^2/4\pi$ and $N_0 = n/\mu$ denote the electron density per spin and
the density of states per spin, respectively.
To obtain Eqs.~\eqref{eq:GL_zeeman} and \eqref{eq:K_zeeman},
the momentum summation is replaced by an integration over $\xi$:
$\frac{1}{V_{\mathrm{vol}}}\sum_{\bm{k}} \rightarrow N_0 \int d\xi$.
To derive Eq.~\eqref{eq:K_zeeman},
we subtract and add the diamagnetic contribution in the normal state
to avoid the formal divergence of the integrand~\cite{agd}.
Equation~\eqref{eq:tqdef} becomes
\begin{align}
    a(T_{\bm{q}}) + \left. \frac{n \, Q(T_{\bm{q}})}{4m|\Delta|^2}\right|_{\Delta=0} q^2 =0, \label{eq:tqdef2}
\end{align}
with $a = a_{\mathrm{z}}$ and $Q=Q_{\mathrm{z}}$ in the present case.
In the Meissner kernel in Eq.~\eqref{eq:K_zeeman}, $Q_{\mathrm{z}}$ is called superfluid weight.
The last term in the numerator of Eq.~\eqref{eq:Q_zeeman}, which stems from the last term in 
Eq.~\eqref{eq:F-func_zeeman}, represents the negative contribution of odd-frequency Cooper pairs to $Q_{\mathrm{z}}$.
Therefore, Eq.~\eqref{eq:tqdef2} has solutions with $q \neq 0$
when the superfluid weight changes the sign for large Zeeman fields~\cite{TS:prb2024}. 

To examine the above argument based on the theory in Sec.~\ref{sec:propagator},
we solve $[D_{\mathrm{s}}^{\mathrm{R}}(\bm{q},0)]^{-1}=0$ in Eq.~\eqref{eq:ds_def} with Eq.~\eqref{eq:pi_def} 
fully numerically for the tight-binding model with $\bm{q} = (q,0)$.
The solution on the $h$--$T$ plane is shown in Fig.~\ref{fig:boundary_q}(a).
Since $[D_{\mathrm{s}}^{\mathrm{R}}(\bm{q},0)]^{-1}=0$ is a single equation for $T_{\bm{q}}$ and $\bm{q}$, 
we obtain a set of solutions $(T_{\bm q}, \bm q)$. 
The highest transition temperatures are plotted in Fig.~\ref{fig:boundary_q}(a) and the corresponding 
$q$ are plotted in (d).
The vertical axis in (a) is normalized to $T_c$, the transition temperature at $h=0$, and
the horizontal axis is normalized to $\Delta_0 \coloneqq \pi e^{-C} T_c = 1.76 T_c$, 
which is the amplitude of the pair potential at $h=T=0$, with $C=0.577$ being Euler's constant~\cite{tinkham}. 
In the numerical simulation, we choose $\mu = t$ and $T_c = 0.05t$.
For weak Zeeman fields, the transition to the uniform state with $q=0$ occurs 
as shown with black dots in Figs.~\ref{fig:boundary_q}(a) and (d).
When a Zeeman field goes beyond a critical value, finite-momentum superconducting states
appear below $T_q$ as shown with the red dots. 
The dependence of $q$ on $h$ in (d) shows that $q$ increases monotonically with $h$.
The transition temperature $T_q$ decreases with increasing $h$ 
and vanishes at $h \sim \Delta_0$.
The phase boundary in Fig.~\ref{fig:boundary_q}(a) slightly deviates from
that for a circular Fermi surface~\cite{matsuda:jpsj2007,burkhardt:annphys1994}
because the direction $\bm{q} = (q,0)$ satisfies a favorable nesting condition
on the anisotropic Fermi surface~\cite{shimahara:prb1994,shimahara:jpsj1997,yokoyama:jpsj2008}.
The results in (d) show that $q$ is much smaller than 
the Fermi wavenumber $k_{\mathrm{F}} \sim 1/c_0$. 
This result, $(qc_0)^2 \ll 1$, justifies the expansion in Eq.~\eqref{eq:DsR}.

The emergence of the FFLO state is usually considered as a result of 
the destruction of Cooper pairs composed of Kramers partners by TRS-breaking perturbations. 
This interpretation provides a clear physical picture of the phenomenon.
However, Eqs.~\eqref{eq:Q_zeeman} and \eqref{eq:tqdef2} suggest an alternative understanding of the FFLO state.
In this picture, the uniform superconducting state is unstable 
because odd-frequency Cooper pairs decrease the superfluid density.
This perspective can also explain the appearance of 
the FFLO states driven by vector potentials~\cite{shimahara:prb1996}, 
as discussed at the end of Sec.~\ref{sec:odd}.
Moreover, finite-momentum superconducting states can appear 
in TRS-preserving SCs because TRS-breaking fields are not essential for 
the emergence of odd-frequency Cooper pairs, as demonstrated in Sec.~\ref{sec:odd}.  
In the next section, we discuss two examples of such SCs.

\section{TRS-preserving finite-momentum states}
\label{sec:FFLO_multiband}
In this section, we discuss two examples of finite-momentum superconducting states in TRS-preserving SCs.
One is a two-band SC with interband pairing order, where 
odd-frequency pairing correlations are induced by band-hybridization and/or band-asymmetry. 
The other is a $j=3/2$ SC with pseudospin-quintet pairing order, where
spin-orbit interactions generate odd-frequency pairs.

\subsection{Two-band SCs with interband pairing orders}
\label{sec:interband}
We consider a two-band SC on a two-dimensional tight-binding square lattice.
In Eq.~\eqref{eq:h}, we choose
\begin{align}
    \hat{\xi}_{\bm{k}} &= \xi_{\bm{k}} \hat{\rho}_0 \hat{\sigma}_0
    + \varepsilon \hat{\rho}_3 \hat{\sigma}_0 + V \hat{\rho}_1 \hat{\sigma}_0 , \\
    \xi_{\bm{k}} &= -2t(\cos k_x + \cos k_y) + 4t - \mu , \label{eq:xik_twoband}\\
    \hat{w}(\bm{k}) 
    &=
    \begin{cases}
        \frac{1}{\sqrt{2}} (i\hat{\rho}_2) \hat{\sigma}_1 & (s=+1) \\
        \frac{1}{\sqrt{2}} \hat{\rho}_1(i\hat{\sigma}_2)  & (s=-1)
    \end{cases} . \label{eq:w_2band}
\end{align}
The length is measured in units of the lattice constant $c_0=1$.
The asymmetry and the hybridization between the two bands are 
represented by $\varepsilon$ and $V$, respectively. 
The pair potential matrix with $s$-wave symmetry is given by
\begin{align}
    &\hat{\Delta}
    =
    \begin{cases}
        \Delta (i\hat{\rho}_2) \hat{\sigma}_1 & (s=+1) \\
        \Delta \hat{\rho}_1(i\hat{\sigma}_2)  & (s=-1)
    \end{cases} , \label{eq:Delta_2band}
\end{align}
where $s=+1(-1)$ represents the odd-band-parity spin-triplet
(even-band-parity spin-singlet) pair potential.
The $8 \times 8$ BdG Hamiltonian can be block-diagonalized
and reduced to two $4 \times 4$ Hamiltonians because 
the normal state Hamiltonian in Eq.~\eqref{eq:xik_twoband} does not 
include any spin-dependent potentials.
One of the $4 \times 4$ Hamiltonians is given by
\begin{align}
    &\check{H}(\bm{k}) =
    \left[
        \begin{array}{cc}
            \hat{H}_{\mathrm{N}} & \hat{\Delta} \\
            \hat{\Delta}^{\dag} & -\hat{H}_{\mathrm{N}} 
        \end{array}
    \right] , \\
    &\hat{H}_{\mathrm{N}} =
    \xi_{\bm{k}} \hat{\rho}_0 + \varepsilon \hat{\rho}_3 + V \hat{\rho}_1 , \\
    &\hat{\Delta} = 
    \begin{cases}
        \Delta (i\hat{\rho}_2) & (s=+1) \\
        \Delta \hat{\rho}_1    & (s=-1)
    \end{cases} .
    \label{eq:H_reduced}
\end{align}

We first discuss the results for odd-band-parity spin-triplet SCs $(s=+1)$. 
It has already been reported that the transition to the uniform superconducting phase becomes discontinuous
for sufficiently large $V$ or $\varepsilon$~\cite{silva:physletta2014,TS:prb2024}.
The anomalous Green's function in the reduced subspace
for the uniform superconducting state is calculated as~\cite{TS:prb2024}
\begin{align}
    &\hat{\mathcal{F}}_{+}(\bm{k}, i\omega_{\ell}) =
    \frac{-1}{Z_{+}} \left[ \omega_{\ell}^2 + 
    \xi_{\bm{k}}^2 - \varepsilon^2 - V^2 + \Delta^2 
    \right. \nonumber \\
       &\hspace{10em}\left.
    - 2i\omega_{\ell} (\varepsilon \hat{\rho}_3 + V \hat{\rho}_1) \right]
    \Delta (i\hat{\rho}_2) ,\label{eq:f_plus} \\
    &Z_{+}
    = \xi_{\bm{k}}^4 + 2\, \xi_{\bm{k}}^2\, A_{+} + C_{+}
    = Z_{\mathrm{z}}|_{h \rightarrow \sqrt{\varepsilon^2 + V^2}} , \label{eq:Z2band+1} \\ 
    &A_{+} = \omega_{\ell}^2 - \varepsilon^2 - V^2 + \Delta^2 , \\
    &C_{+} = A^2_{+} + 4 \omega_{\ell}^2\, (\varepsilon^2 + V^2) .
\end{align}
Both the band hybridization $V$ and the band asymmetry $\varepsilon$ induce 
pairing correlations belonging to the odd-frequency symmetry class,
as shown in the last term of Eq.~\eqref{eq:f_plus}.
The BdG Hamiltonian is equivalent to that of a conventional 
spin-singlet SC under Zeeman fields in Sec.~\ref{sec:FFLO}.
The GL coefficient and the Meissner kernel
have the same structure as those in Eqs.~\eqref{eq:GL_zeeman}, \eqref{eq:K_zeeman} and \eqref{eq:Q_zeeman}.
In the continuum limit, the results are given by
\begin{align}
    &a_{+}(T) = \frac{1}{g} - 2N_0 \pi T \sum_{\omega_{\ell}} 
    \frac{\omega_{\ell}^2}{|\omega_{\ell}| (\omega_{\ell}^2 + \varepsilon^2 + V^2)} , \label{eq:GL_2band+1} \\
    &K_{+,\mu\nu} (\bm{0},0) = \delta_{\mu\nu} \frac{4ne^2}{m}
    Q_{+} , \label{eq:K_2band+1} \\
    &Q_{+} = Q_{\mathrm{z}}|_{h \rightarrow \sqrt{\varepsilon^2+V^2}} , \label{eq:Q_2band+1}
\end{align}
where $Q_{\mathrm{z}}$ is defined in Eq.~\eqref{eq:Q_zeeman}.
Therefore, $K_{+,\mu\nu}$ changes sign and becomes negative
for sufficiently large $\sqrt{\varepsilon^2 + V^2}$~\cite{TS:prb2024}.
In the numerical simulation, we use the same parameters as those in Sec.~\ref{sec:FFLO}: 
$\mu=t$, $T_c=0.05t$, $\Delta_0 = \pi e^{-C} T_c$, and $\bm{q}=(q,0)$.
The phase boundary for $s=+1$ can be obtained by replacing the horizontal axis $h$ with 
$\sqrt{\varepsilon^2 + V^2}$ in Figs.~\ref{fig:boundary_q}(a) and (d). 

For even-band-parity spin-singlet SCs with $s=-1$, we obtain
\begin{align}
    &\hat{\mathcal{F}}_{-}(\bm{k}, i\omega_{\ell}) =
    \frac{-1}{Z_{-}} \left[ \omega_{\ell}^2 + \xi_{\bm{k}}^2 - \varepsilon^2 + V^2 + \Delta^2 \right. \nonumber \\
    &\hspace{6em}\left.
    - 2(\xi_{\bm{k}} V\hat{\rho}_1 - i\varepsilon V \hat{\rho}_2 + i\omega_{\ell} \varepsilon\hat{\rho}_3 ) \right]
    \Delta \hat{\rho}_1 , \\
    &Z_{-}
    = \xi_{\bm{k}}^4 + 2\, \xi_{\bm{k}}^2\, A_{-} + C_{-} , \label{eq:Z2band-1} \\
    &A_{-} = A_{+} , \\
    &C_{-} = A^2_{-} + 4 \left(\omega_{\ell}^2 \varepsilon^2 + \omega_{\ell}^2 V^2 + V^2 \Delta^2 \right).
\end{align}
The band hybridization $V$ induces subdominant even-frequency pairing correlation, whereas
the band asymmetry $\varepsilon$ induces both even- and odd-frequency pairing correlations.
The GL coefficient and the Meissner kernel in the continuum limit are given by 
\begin{align}
    &a_{-}(T) = \frac{1}{g} - 2N_0 \pi T \sum_{\omega_{\ell}} 
    \frac{\omega_{\ell}^2 + V^2}{|\omega_{\ell}| (\omega_{\ell}^2 + \varepsilon^2 + V^2)} , \label{eq:GL_2band-1} \\
    &K_{-,\mu\nu} (\bm{0},0) = \delta_{\mu\nu} \frac{4ne^2}{m}
    Q_{-} , \label{eq:K_2band-1} \\
    &Q_{-} = \sqrt{2} \pi T \sum_{\omega_{\ell}}
    \frac{\Delta^2}
    {[ C_{-} (A_{-}+\sqrt{C_{-}})]^{3/2}} \nonumber \\
    &\times\left[ A_{-}^3+\sqrt{C_{-}}(A_{-}^2-2\omega_{\ell}^2 \varepsilon^2) \right. \nonumber \\
    &\hspace{1em}\left.+2V^2 \left\{ C_{-} + 2(\omega_{\ell}^2+\Delta^2) (A_{-}+\sqrt{C_{-}}) \right\}\right]
    . \label{eq:Q_2band-1}
\end{align}
While the odd-frequency correlation induced by the band 
asymmetry $\varepsilon$
reduces $K_{-,\mu\nu}$ and eventually changes its sign to negative, 
the even-frequency correlation induced by the hybridization $V$ enhances the Meissner kernel.
Thus, finite-momentum superconducting states are expected for sufficiently large $\varepsilon$.
To confirm this, we numerically solve $[D_{\mathrm{s}}^{\mathrm{R}}(\bm{q},0)]^{-1}=0$ 
on the tight-binding model.
The parameters in the numerical simulation are the same as those
in Sec.~\ref{sec:FFLO}: $\mu=t$, $T_c=0.05t$, $\Delta_0 = \pi e^{-C} T_c$, and $\bm{q}=(q,0)$.
The results for $V=0.3\Delta_0$ are shown in Fig.~\ref{fig:boundary_q}(b).
The magnitude of $\bm{q}$ at the transition point is also plotted as a function of $\varepsilon$ in (e). 
The transition to the finite-momentum superconducting state occurs for $\varepsilon \gtrsim 0.6 \Delta_0$,
as indicated by red dots in (b) and (e). 
The characteristic features in (b) and (e) are essentially the same as those in (a) and (d).
The resulting $q$ are much smaller than $k_{\mathrm{F}} \sim 1/c_0$, which 
justifies the expansion in Eq.~\eqref{eq:DsR}.

\subsection{$j=3/2$ SC with a pseudospin-quintet pairing order}
\label{sec:j32}
An electron in a $j=3/2$ SC
is characterized by a total angular momentum $j=3/2$, resulting from strong coupling between 
spin $s=1/2$ and orbital angular momentum $\ell=1$~\cite{brydon:prl2016}.  
The $s$-wave pseudospin-quintet pair potential 
represents a superconducting state due to the Cooper pairing between 
two such high-pseudospin electrons.
To describe the electronic structure on a simple cubic lattice, we choose
\begin{align}
    \hat{\xi}_{\bm{k}} &=
    \xi_{\bm{k}} 1_{4\times 4} + \vec{\epsilon}_{\bm{k}} \cdot \vec{\gamma} 
    \label{eq:normal_j32} , \quad
    \hat{w}(\bm{k}) = \frac{1}{\sqrt{2}} \gamma^4 U_T , \\
    \xi_{\bm{k}} &= \left( -2 t_1 - \frac{5}{2} t_2 \right) \sum_{\nu} \cos{k_{\nu}} + 6 t_1 + \frac{15}{2} t_2 - \mu , \\
    \epsilon_{\bm{k},1} &= 4\sqrt{3} t_3 \sin{k_x} \sin{k_y} , \\
    \epsilon_{\bm{k},2} &= 4\sqrt{3} t_3 \sin{k_y} \sin{k_z} , \\
    \epsilon_{\bm{k},3} &= 4\sqrt{3} t_3 \sin{k_z} \sin{k_x} , \\
    \epsilon_{\bm{k},4} &= \sqrt{3} t_2 ( -\cos{k_x} + \cos{k_y} ) , \\
    \epsilon_{\bm{k},5} &= t_2 ( -2 \cos{k_z} + \cos{k_x} + \cos{k_y} ) , \\
    \label{eq:hn_gamma}
    U_T &= \gamma^1 \gamma^2 ,
\end{align}
in Eq.~\eqref{eq:h}~\cite{luttinger:pr1955,sigrist:rmp1991,brydon:prl2016,agterberg:prl2017,brydon:prb2018,roy:prb2019}, 
where the definitions of the five $4\times 4$ matrices $\gamma^{\mathrm{j}}$ for $\mathrm{j}=1-5$ 
and their algebra are shown in Appendix~\ref{sec:algebras}.
The length is measured in units of the lattice constant $c_0=1$.
The dispersion $\xi_{\bm{k}}$ is independent of the pseudospin of an electron, 
whereas that of the five-component vector
$\vec{\epsilon}_{\bm{k}} = (\epsilon_{\bm{k},1}, \epsilon_{\bm{k},2}, \epsilon_{\bm{k},3}, \epsilon_{\bm{k},4}, \epsilon_{\bm{k},5})$
depends on pseudospin.
The pair potential is given by $\hat{\Delta}(\bm{k}) = \Delta \gamma^4 U_T$.
Although we set $t_2=0$ for simplicity, we retain $t_3 \ll t_1$ as a source of odd-frequency pairing correlations.
The anomalous Green's function for the uniform superconducting state is given by~\cite{TS:prb2024}
\begin{align}
    &\hat{\mathcal{F}}_{3/2}(\bm{k},i\omega_{\ell}) =
    -\frac{\Delta}{Z_{3/2}}
    \left[
        W -2i\omega_{\ell} \vec{\epsilon}_{\bm{k}} \cdot \vec{\gamma}
    \right]
    \gamma^4 U_{T} , \label{eq:F-func_j32}\\
    &Z_{3/2} = W^2 + 4\omega_{\ell}^2 \vec{\epsilon}_{\bm{k}}^{\,2}, \quad
    W = \omega_{\ell}^2 + \xi_{\bm{k}}^2 - \vec{\epsilon}_{\bm{k}}^{\,2}  + \Delta^2 \label{eq:zw_main} .
\end{align}
The spin-orbit interaction $\vec{\epsilon}_{\bm{k}}$ induces
a subdominant odd-frequency pairing correlation, as shown in the second term 
in Eq.~\eqref{eq:F-func_j32}~\cite{TS:prb2024}. 
The GL coefficient and the contribution of the anomalous Green's function
to the Meissner kernel, $K^{\mathcal{F}}$, in the lattice model are given by~\cite{kim:jpsj2021,dutta:prr2021,TS:prb2024} 
\begin{align}
    &a_{3/2}(T) =
    \frac{1}{g}
    -2T\sum_{\omega_{\ell}} \frac{1}{N}\sum_{\bm{k}}
    \frac{1}{Z_{3/2}|_{\Delta=0}}
    ( \omega_{\ell}^2 + \xi_{\bm{k}}^2 - \vec{\epsilon}_{\bm{k}}^{\,2} )
    \label{eq:GL_j32} , \\
    &K^{\mathcal{F}}_{3/2,xx} (\bm{0},0) = \nonumber \\
    &\hspace{1em} 16 e^2 T\sum_{\omega_{\ell}} \frac{1}{N}\sum_{\bm{k}}
    t^2_1 \sin^2 k_x \frac{\Delta^2}{Z_{3/2}^2}\left[
    W^2 - 4 \omega_{\ell}^2 \, \vec{\epsilon}_{\bm{k}}^{\,2}
    \right] \label{eq:KF_j32} ,
\end{align}
where, for simplicity, we neglect the correction to the velocity operator arising
from the weak spin-orbit interaction $(t_3 \ll t_1)$~\cite{TS:prb2024}.
The second term in Eq.~\eqref{eq:KF_j32} represents the negative contribution
of odd-frequency correlations to the Meissner kernel.
Therefore, finite-momentum superconducting states can be expected 
when the amplitude of the spin-orbit interaction $t_3$ is sufficiently large.
Figures~\ref{fig:boundary_q}(c) and (f) show the numerical 
solutions of $[D_{\mathrm{s}}^{\mathrm{R}}(\bm{q},0)]^{-1}=0$, 
where we choose $\mu = t_1$, $t_2=0$, $T_c = 0.05t_1$, $\Delta_0 = \pi e^{-C} T_c$, and $\bm{q} = (q,0,0)$.
The characteristic features in (c) and (f) are qualitatively the same as those in (a) and (d).
The resulting $q$ are much smaller than $k_{\mathrm{F}} \sim 1/c_0$, which 
justifies the expansion in Eq.~\eqref{eq:DsR}.

\section{Discussion}
\label{sec:discussion}
\subsection{Platforms for finite-momentum superconductivity}
\label{sec:candidate}
In Sec.~\ref{sec:FFLO_multiband}, 
we demonstrate the emergence of finite-momentum superconductivity in TRS-preserving SCs. 
Here, we briefly discuss the possibility in real materials.
In the present analysis, we focus only on interband/orbital superconductivity, 
in which electrons belonging to different bands or orbitals experience attractive interactions, 
and do not consider intraband/orbital superconductivity, such as that in 
MgB$_2$~\cite{mgb2:akimitsu2001,mgb2:louie2002} or 
iron-based compounds~\cite{pnictide:hosono2008,kuroki:prl2008}. 
This is because the magnitudes of odd-frequency pairing correlations in such
intraband/orbital SCs tend to be small~\cite{sasaki:prb2020}. 
Thus, the instability of the uniform state is not expected. 

Doped topological insulator Cu$_x$Bi$_2$Se$_3$ and related compounds~\cite{fu:prl2010,matano:natphys2016}
can be representative examples of interband/orbital SCs. 
In addition, topological superconductivity proposed in bilayer Rashba systems~\cite{nakosai:prl2012} 
provides another platform for realizing such pairing states.
Although the emergence of odd-frequency pairing correlations 
in Cu$_x$Bi$_2$Se$_3$ has been discussed~\cite{TS:prb2020,mizushima:prb2023}, 
finite-momentum superconducting states have not been reported.
Another example is transition-metal dichalcogenide (TMD) SCs, where two valleys 
located at $K$ and $K^\prime$ points in the Brillouin zone correspond to 
an extra internal degree of freedom of an electron~\cite{saito:natphys2016, xi:natphys2016, barrera:natcomm2018}.
The pair potential for the Kramers partner is described as an intervalley pair potential. 
The Hamiltonian of a $j = 3/2$ SC~\cite{brydon:prl2016} can capture the low-energy 
electronic structure of a wide class of two-band/orbital SCs, 
including the above interband/orbital superconducting systems~\cite{cavanagh:prb2023}. 
Therefore, TRS-preserving finite-momentum superconducting states 
may be realized in these materials.

\subsection{Relation to the discontinuous transition to a uniform state}
\label{sec:discontinuous}
It is known that the following relation holds between two transition temperatures 
of a conventional SC in Zeeman fields~\cite{matsuda:jpsj2007}. 
\begin{align}
T_{\mathrm{dis}} = T_{\mathrm{FFLO}}, \label{eq:tt}
\end{align}
where $T_{\mathrm{FFLO}}$ is the highest second-order transition temperature to the finite-momentum 
state in the $h$--$T$ phase diagram, and $T_{\mathrm{dis}}$ is the highest first-order (discontinuous) transition temperature 
to the uniform superconducting state~\cite{sarma:jpcs1963,maki_tsuneto:ptp1964}.
The reason for the coincidence is well explained by the relations
among the superfluid weight $Q(T)$, the quartic coefficient $b(T)$ in the GL
free-energy for a uniform SC in Eq.~\eqref{eq:Omega_GL}, and 
the coefficient $B(T)$ in Eq.~\eqref{eq:DsR}.
This study shows that the relation
\begin{align}
   B(T) = A_B\,  \left. Q(T) / |\Delta|^2 \right|_{\Delta\to 0}, \label{eq:bq1} 
\end{align}
holds with a constant $A_B>0$, and that $T_{\mathrm{FFLO}}$ is well characterized by $B=Q=0$.
In a previous study~\cite{TS:prb2024},
we also found that the relation
\begin{align}
   b(T)\approx A_b\,  \left. Q(T) / |\Delta|^2 \right|_{\Delta\to 0}, \label{eq:bq-relation} 
\end{align}
is satisfied near the transition temperature for uniform SCs, with $A_b>0$ being a constant,
and that $T_{\mathrm{dis}}$ is characterized by $b=Q=0$%
~\footnote{
    It is not possible to carry out the integration over momenta analytically in Eq.~\eqref{eq:K(0,0)}
	in many cases. 
    In such cases, $Q$ in Eq.~\eqref{eq:bq-relation} can be replaced by $Q^{\mathcal{F}}$, 
    which corresponds to the contribution of the anomalous Green's function to the superfluid density.
}.
In such cases, the sign change of the superfluid weight $Q$ 
characterizes both $T_{\mathrm{dis}}$ and $T_{\mathrm{FFLO}}$ simultaneously,
thereby explaining their coincidence.
Equation~\eqref{eq:tt} holds as long as the relation in Eq.~\eqref{eq:bq-relation} is satisfied
and the higher order terms in Eq.~\eqref{eq:DsR} are negligible
\footnote{
    Of course, this argument is not valid
    when the electron correlations in the normal state are very strong~\cite{burkhardt:annphys1994}.
    However, the relation is expected to hold in most weak-coupling SCs.
}.
It should be noted that the relation between the sign of the superfluid density and the stability of finite-momentum 
superconducting states was also implicit in a previous study~\cite{taylor:pra2006}.

\section{Conclusion}
\label{sec:conclusion}
We have discussed the mechanisms for the emergence of inhomogeneous superconductivity 
characterized by a small center-of-mass momentum $q$ of Cooper pairs.
The analytic expression for the pair fluctuation propagator provides the conditions 
for the appearance of such finite-momentum superconductivity. 
When the superfluid density expected in the uniform superconducting state is negative,
a finite-momentum superconducting state appears below the transition temperature. 
The analytic expressions for the superfluid density indicate that the 
odd-frequency Cooper pairing correlations reduce the superfluid density in 
a conventional SC in Zeeman fields, two-band SCs with interband pairing order, and 
a $j=3/2$ SC. Time-reversal symmetry is preserved in the latter two cases.
In all cases, we demonstrate the transition to finite-momentum superconducting states 
using both the analytic expression for the pair fluctuation propagator and numerical calculations.
We conclude that the instability of the uniform superconducting state  
due to odd-frequency pairing correlations is a source of the emergence of finite-momentum superconducting states.

We have developed a theoretical framework that explains the 
emergence of finite-momentum superconducting states by considering the negative 
superfluid density induced by odd-frequency pairing correlations.
Our theory provides a unified understanding of inhomogeneous superconducting states in bulk%
~\cite{fulde_fflo,larkin_fflo,mineev:prb2008,li:prb2024} 
and around local defects%
~\cite{tanuma:prl2009,kuzmanovski:prb2020,perrin:prl2020,shu:prb2022,shu:jpscp2023,shu:prb2023,tanaka:prl2007,asano:prb2013}.

\section*{Acknowledgments}
T. S. is grateful to S.~Ikegaya and K.~Aoyama for useful discussions.
S. K. was supported by JSPS KAKENHI (Grants No. JP19K14612 and No. JP22K03478) and JST CREST (Grant No. JPMJCR19T2).
S. H. was supported by JSPS KAKENHI (Grants No. JP21H01037 and No. JP23H04869) and JST FOREST (Grant No. JPMJFR2366).
Y. A. was supported by a Grant-in-Aid for Scientific Research (JSPS KAKENHI Grant No. JP26K0692).

\appendix
\section{Meissner kernel up to $\Delta^2$}
\label{sec:meissner}
To discuss the linear response of a uniform SC described by $\mathcal{H}$ in Eq.~\eqref{eq:h} to an external magnetic field,
we express the noninteracting Hamiltonian in Eq.~\eqref{eq:h0k} in real space:
\begin{align}
    \mathcal{H}_0 &=
    \sum_{\bm{r}\bm{R}\alpha\beta} \left( -t_{\bm{R}}^{\alpha\beta} -\mu \delta_{\bm{R},\bm{0}} \delta_{\alpha\beta} \right)
    c^\dag_{\bm{r}+\bm{R}\alpha} c_{\bm{r}\beta} .
\end{align}
The matrix element of the noninteracting Hamiltonian in momentum space is given by
$\epsilon_{\bm{k}\alpha\beta} = \sum_{\bm{R}} -t^{\alpha\beta}_{\bm{R}} e^{-i\bm{k}\cdot\bm{R}}$.
The coupling between electrons and an electromagnetic field is introduced through the Peierls phase
in the hopping $t^{\alpha\beta}_{\bm{R}}$ in $\mathcal{H}_0$~\cite{peierls:zphys1933,luttinger:pr1951}:
\begin{align}
    -t^{\alpha\beta}_{\bm{R}} &\rightarrow -t^{\alpha\beta}_{\bm{R}} e^{i\varphi_{\bm{R}}(\bm{r})} , \\
    \varphi_{\bm{R}}(\bm{r}) &= e\int_{\bm{r}}^{\bm{r}+\bm{R}} d\bm{r} \cdot \bm{A}(\bm{r})
    \approx e\bm{A}(\bm{r}+\bm{R}/2) \cdot \bm{R} ,
    \label{eq:phiapprox}
\end{align}
where the approximation in Eq.~\eqref{eq:phiapprox} is justified
when the vector potential varies on a length scale much larger than the lattice spacing.
We also assume that the Peierls phase is independent of the internal degrees of freedom of an electron 
and that the vector potential does not couple to the effective electron-electron 
interaction Hamiltonian $\mathcal{H}_1$.
The current density operator is defined via the variation of the Hamiltonian with respect to the vector potential:
\begin{align}
    \delta \mathcal{H}_{\bm{A}} &= \mathcal{H}_{\bm{A}+\delta \bm{A}} - \mathcal{H}_{\bm{A}}
    = -\sum_{\bm{\ell}} \bm{j}_{\bm{\ell}} \cdot \delta \bm{A}_{\bm{\ell}} + O(\delta A^2) ,
\end{align}
where $\mathcal{H}_{\bm{A}}$ is the total Hamiltonian including the Peierls phase.
The current density operator is then given by 
\begin{align}
    \bm{j}_{\bm{\ell}} (t) \coloneqq \sum_{\substack{\bm{r}\bm{R} \\ \alpha\beta}} ie\bm{R} \ t^{\alpha\beta}_{\bm{R}} \
    e^{ie\bm{A}_{\bm{\ell}}\cdot\bm{R}} \ c^{\dag}_{\bm{r}+\bm{R}\alpha} c_{\bm{r}\beta} \ \delta_{\bm{\ell},\bm{r}+\bm{R}/2} .
\end{align}
To first order in the vector potential, the current density operator can be decomposed into paramagnetic and diamagnetic contributions: 
\begin{align}
    j_{\mu\bm{\ell}} (t) &= j^{\mathrm{para}}_{\mu\bm{\ell}} + j^{\mathrm{dia}}_{\mu\bm{\ell}}(t) + O(A^2) , \\
    j^{\mathrm{para}}_{\mu\bm{\ell}} &\coloneqq ie\sum_{\substack{\bm{r}\bm{R} \\ \alpha\beta}} R_{\mu} \ t^{\alpha\beta}_{\bm{R}} \ 
    c^{\dag}_{\bm{r}+\bm{R}\alpha} c_{\bm{r}\beta} \ \delta_{\bm{\ell},\bm{r}+\bm{R}/2} , \\
    j^{\mathrm{dia}}_{\mu\bm{\ell}}(t) &\coloneqq -e^2\sum_{\substack{\bm{r}\bm{R} \\ \alpha\beta}} R_{\mu}R_{\nu} \ t^{\alpha\beta}_{\bm{R}} \ 
    c^{\dag}_{\bm{r}+\bm{R}\alpha} c_{\bm{r}\beta} \ A_{\nu\bm{\ell}}(t) \ \delta_{\bm{\ell},\bm{r}+\bm{R}/2} .
\end{align}
The total Hamiltonian in the presence of a vector potential reads
\begin{align}
    \mathcal{H}_{\bm{A}} = \mathcal{H} - \sum_{\bm{\ell}} \bm{j}^{\mathrm{para}}_{\bm{\ell}} \cdot \bm{A}_{\bm{\ell}}(t) + O(A^2) .
\end{align}
We consider only transverse gauge fields (i.e., $\bm{q} \cdot \bm{A}_{\bm{q}} = 0$).
The expectation value of the current density is calculated
using linear response theory~\cite{scalapino:prl1992,scalapino:prb1993,kostyrko:prb1994}:
\begin{align}
    \langle j_{\mu\bm{q}} \rangle (\omega)
    &= -K_{\mu\nu} (\bm{q},\omega) A_{\nu\bm{q}} (\omega) \nonumber \\
    &= \langle j_{\mu\bm{q}} (\omega) \rangle
    + \Lambda^{\mathrm{R}}_{\mu\nu} (\bm{q},\omega) A_{\nu\bm{q}}(\omega) ,
\end{align}
where $\langle j_{\mu\bm{q}} (\omega) \rangle$ and $\Lambda^{\mathrm{R}}_{\mu\nu} (\bm{q},\omega)$ are the Fourier components
of the expectation value of the current density operator and the current-current correlation function, respectively.
Within the mean-field approximation, they are calculated as follows: 
\begin{widetext}
\begin{align}
    \langle j_{\mu\bm{q}} (\omega) \rangle
    &= -e^2 T\sum_{\omega_{\ell}} \frac{1}{N} \sum_{\bm{k}}
    \mathrm{Tr} \left[ \partial_{k_{\mu}} \partial_{k_{\nu}} \hat{\epsilon}_{\bm{k}} \hat{\mathcal{G}} (\bm{k},i\omega_{\ell}) \right]
    e^{i\omega_{\ell} \eta} A_{\nu \bm{q}} (\omega) , \\
    \Lambda^{\mathrm{R}}_{\mu\nu} (\bm{q},\omega) &= \Lambda_{\mu\nu} (\bm{q},i\nu_n \rightarrow \omega+i\delta) , \\
    \Lambda_{\mu\nu} (\bm{q},i\nu_n) &= \int_{0}^{\beta} d\tau \ 
    \frac{1}{N} \langle T_{\tau} j^{\mathrm{para}}_{\mu\bm{q}}(\tau) j^{\mathrm{para}}_{\nu -\bm{q}} \rangle \ e^{i\nu_n \tau} \\
    &= -e^2 T\sum_{\omega_{\ell}} \frac{1}{N} \sum_{\bm{k}} \mathrm{Tr}
    \left[
        \hat{v}_{\mu}(\bm{k}+\bm{q}/2)\hat{\mathcal{G}}(\bm{k}+\bm{q},i\omega_{\ell}+i\nu_n)
        \hat{v}_{\nu}(\bm{k}+\bm{q}/2)\hat{\mathcal{G}}(\bm{k},i\omega_{\ell})
    \right. \nonumber \\
    &\hspace{9em}\left.
        +\hat{v}_{\mu}(\bm{k}+\bm{q}/2)\hat{\mathcal{F}}(\bm{k}+\bm{q},i\omega_{\ell}+i\nu_n)
        \hat{\underline{v}}_{\nu}(\bm{k}+\bm{q}/2)\hat{\mathcal{\underline{F}}}(\bm{k},i\omega_{\ell})
    \right] ,
\end{align}
where we assumed $\langle j^{\mathrm{para}}_{\mu\bm{q}} \rangle = 0$, $\eta$ and $\delta$ are small positive real values, and
$\mathcal{G} \, (\mathcal{F})$ is the normal (anomalous) Green's function. 
The mean-field Hamiltonian and the Green's functions are given by
\begin{align}
    &\mathcal{H}^{\mathrm{MF}} = \mathcal{H}_0 + \mathcal{H}^{\mathrm{MF}}_1
    = \frac{1}{2} \sum_{\bm{k}} \vec{\Psi}^{\dag}_{\bm{k}} H(\bm{k}) \vec{\Psi}_{\bm{k}} + \mathrm{const.}
    \label{eq:HMF}, \\
    &H(\bm{k}) = \left[
        \begin{array}{cc}
            \hat{\xi}_{\bm{k}} & \hat{\Delta}(\bm{k}) \\
            -\hat{\Delta}^{\ast}(-\bm{k}) & -\hat{\xi}^{\ast}_{-\bm{k}}
        \end{array}
    \right] , \quad
    \vec{\Psi}_{\bm{k}} = \left[ \vec{\psi}^{\,\mathrm{T}}_{\bm{k}}, \vec{\psi}^{\,\mathrm{\dag}}_{-\bm{k}} \right]^{\,\mathrm{T}} , \quad
    \vec{\psi}_{\bm{k}}  = \left[ c_{\bm{k}1}, c_{\bm{k}2}, \cdots, c_{\bm{k}M} \right]^{\mathrm{T}} 
    \label{eq:HBdG}, \\
    &\left[
        \begin{array}{cc}
            \hat{\mathcal{G}} & \hat{\mathcal{F}} \\
            \hat{\underline{\mathcal{F}}} & \hat{\underline{\mathcal{G}}}
        \end{array}
    \right]_{(\bm{k},i\omega_{\ell})} = [i\omega_{\ell} - H(\bm{k})]^{-1} 
    \label{eq:gorkov},
\end{align}
where $M$ is the number of internal degrees of freedom of the electron,
$H(\bm{k})$ is a $2M \times 2M$ matrix, and the mean field is defined as follows:
\begin{align}
    \Delta_{\sigma\sigma'} (\bm{k}) &\coloneqq
    w_{\sigma \sigma'}(\bm{k}) \sum_{\gamma\delta} \frac{g}{N} \sum_{\bm{k}'} w^{\ast}_{\gamma\delta}(\bm{k}')
    \langle c_{\bm{k}'\gamma} c_{-\bm{k}'\delta} \rangle . \label{eq_a:defdelta}
\end{align}
The response to a static, uniform magnetic field is described by $K_{\mu\nu}(\bm{q} \rightarrow \bm{0},\omega=0)$.
We obtain
\begin{align}
    K_{\mu\nu}(\bm{0},0) &=
    e^2 T\sum_{\omega_{\ell}} \frac{1}{N} \sum_{\bm{k}} \mathrm{Tr} \left[
        \partial_{k_{\mu}} \partial_{k_{\nu}} \hat{\epsilon}_{\bm{k}} \ \hat{\mathcal{G}} (\bm{k},i\omega_{\ell}) e^{i\omega_{\ell} \eta}
        \right. \nonumber \\
        &\hspace{8em}\left.
        +\hat{v}_{\mu}(\bm{k})\hat{\mathcal{G}}(\bm{k},i\omega_{\ell})
        \hat{v}_{\nu}(\bm{k})\hat{\mathcal{G}}(\bm{k},i\omega_{\ell})
        +\hat{v}_{\mu}(\bm{k})\hat{\mathcal{F}}(\bm{k},i\omega_{\ell}) 
        \hat{\underline{v}}_{\nu}(\bm{k})\hat{\mathcal{\underline{F}}}(\bm{k},i\omega_{\ell})
    \right]
    \label{eq:K(0,0)} .
\end{align}
The contribution from the anomalous Green's function is defined as
\begin{align}
    K^{\mathcal{F}}_{\mu\nu}(\bm{0},0) \coloneqq
    e^2 T\sum_{\omega_{\ell}} \frac{1}{N} \sum_{\bm{k}} \mathrm{Tr} \left[
        \hat{v}_{\mu}(\bm{k})\hat{\mathcal{F}}(\bm{k},i\omega_{\ell}) 
        \hat{\underline{v}}_{\nu}(\bm{k})\hat{\mathcal{\underline{F}}}(\bm{k},i\omega_{\ell})
    \right] ,
\end{align}
in Eq.~\eqref{eq:KF_j32}.
Expanding the Green's function with respect to $\Delta$, we obtain
\begin{align}
    K_{\mu\nu}(\bm{0},0) = e^2 T\sum_{\omega_{\ell}} \frac{1}{N} \sum_{\bm{k}} \mathrm{Tr}
    &\left[
        \partial_{k_{\mu}} \partial_{k_{\nu}} \hat{\epsilon} \ \hat{\mathcal{G}}_0 \hat{\Delta} \hat{\mathcal{\underline{G}}}_0 \hat{\underline{\Delta}}
        \hat{\mathcal{G}}_0 e^{i\omega_{\ell}\eta} 
        +\hat{v}_{\mu} \hat{\mathcal{G}}_0 \hat{v}_{\nu} \hat{\mathcal{G}}_0 \hat{\Delta} \hat{\mathcal{\underline{G}}}_0 \hat{\underline{\Delta}}
        \hat{\mathcal{G}}_0 \right. \nonumber \\
        &+\left.\hat{v}_{\mu} \hat{\mathcal{G}}_0 \hat{\Delta} \hat{\mathcal{\underline{G}}}_0 \hat{\underline{\Delta}} \hat{\mathcal{G}}_0 \hat{v}_{\nu}
        \hat{\mathcal{G}}_0
        +\hat{v}_{\mu} \hat{\mathcal{G}}_0 \hat{\Delta} \hat{\mathcal{\underline{G}}}_0 \hat{\underline{v}}_{\nu} \hat{\mathcal{\underline{G}}}_0 \hat{\underline{\Delta}}
        \hat{\mathcal{G}}_0
    \right] + O(\Delta^4) \label{eq:Kexpand} ,
\end{align}
where we assume the absence of the Meissner effect in the normal state:
\begin{align}
    K^{\mathrm{N}}_{\mu\nu}(\bm{0},0) \coloneqq e^2 T\sum_{\omega_{\ell}} \frac{1}{N} \sum_{\bm{k}} \mathrm{Tr}
    &\left[
        \partial_{k_{\mu}} \partial_{k_{\nu}} \hat{\epsilon} \ \hat{\mathcal{G}}_0 e^{i\omega_{\ell}\eta} 
        +\hat{v}_{\mu} \hat{\mathcal{G}}_0 \hat{v}_{\nu} \hat{\mathcal{G}}_0 
    \right] =0 .
\end{align}
\end{widetext}

\section{Normal Green's function}
\label{sec:g-func}
The normal Green's functions in the uniform superconducting states
of the three theoretical models are summarized in this section.
The result for a conventional SC under a Zeeman field discussed in Sec.~\ref{sec:FFLO} reads as follows: 
\begin{align}
    &\hspace{1em}\hat{\mathcal{G}}_{\mathrm{z}}(\bm{k}, i\omega_{\ell}) = \nonumber \\
    &\hspace{2em}\frac{-1}{Z_{\mathrm{z}}}
    \left[
    (i\omega_{\ell} + \xi_{\bm{k}}) (\omega_{\ell}^2+\xi_{\bm{k}}^2+\Delta^2) + (i\omega_{\ell} - \xi_{\bm{k}}) h^2
    \right. \nonumber \\
    &\hspace{3em}\left. + 
    \left\{  (i\omega_{\ell} + \xi_{\bm{k}})^2 -h^2 +\Delta^2  \right\} \bm{h} \cdot \hat{\bm{\sigma}}
    \right] .
\end{align}

The result for a two-band SC with $s=+1$ discussed in Sec.~\ref{sec:interband} is given by 
\begin{align}
    &\hat{\mathcal{G}}_{+}(\bm{k}, i\omega_{\ell}) = \nonumber \\
    &\frac{-1}{Z_{+}}
    \left[
    (i\omega_{\ell} + \xi_{\bm{k}}) (\omega_{\ell}^2+\xi_{\bm{k}}^2+\Delta^2) + (i\omega_{\ell} - \xi_{\bm{k}}) (\varepsilon^2 + V^2)
    \right. \nonumber \\
    &\left. -
    \left\{  (i\omega_{\ell} + \xi_{\bm{k}})^2 -(\varepsilon^2+V^2) +\Delta^2 \right\}
    (\varepsilon \hat{\rho}_3 + V \hat{\rho}_1)
    \right] .
\end{align}
For $s=-1$, we obtain 
\begin{align}
    &\hat{\mathcal{G}}_{-}(\bm{k}, i\omega_{\ell}) = \nonumber \\
    &\frac{-1}{Z_{-}}
    \left[
    (i\omega_{\ell} + \xi_{\bm{k}}) (\omega_{\ell}^2+\xi_{\bm{k}}^2+\Delta^2) + (i\omega_{\ell} - \xi_{\bm{k}}) (\varepsilon^2 + V^2)
    \right. \nonumber \\
    &\hspace{2em}-
    \left\{  (i\omega_{\ell} + \xi_{\bm{k}})^2 -(\varepsilon^2+V^2) +\Delta^2 \right\}
    (\varepsilon \hat{\rho}_3 + V \hat{\rho}_1) \nonumber \\
    &\hspace{2em}\left. + 2V\Delta^2 \hat{\rho}_1
    \right] .
\end{align}

In the $j=3/2$ SC discussed in Sec.~\ref{sec:j32} with $t_2=0$, we obtain
\begin{align}
    &\hat{\mathcal{G}}_{3/2}(\bm{k},i\omega_n) = \nonumber \\
    &\frac{-1}{Z_{3/2}}
    \left[
        (i\omega_n + \xi_{\bm{k}}) (\omega_n^2 + \xi_{\bm{k}}^2  + \Delta^2)
        +(i\omega_n - \xi_{\bm{k}}) \vec{\epsilon}_{\bm{k}}^{\,2} \right. \nonumber \\
    &\hspace{2em}\left. -\left\{ (i\omega_n + \xi_{\bm{k}})^2 - \vec{\epsilon}_{\bm{k}}^{\,2} + \Delta^2
        \right\} \vec{\epsilon}_{\bm{k}} \cdot \vec{\gamma}
    \right] . \label{eq:g-func}
\end{align}

\section{Algebras of $\gamma$ matrices}
\label{sec:algebras}
The angular momentum operators for $j=3/2$ electrons are given by
\begin{align}
J_x &= \frac{1}{2}\left[\begin{array}{cccc}
0 & \sqrt{3} & 0 & 0 \\
\sqrt{3} & 0 & 2 & 0\\
0 & 2 & 0 & \sqrt{3} \\
0 & 0 & \sqrt{3} & 0
\end{array}\right], \\
J_y &= \frac{1}{2}\left[\begin{array}{cccc}
0 & -i\sqrt{3} & 0 & 0 \\
i\sqrt{3} & 0 & -2i & 0\\
0 & 2i & 0 & -i\sqrt{3} \\
0 & 0 & i\sqrt{3} & 0
\end{array}\right], \\
J_z &= \frac{1}{2}\left[\begin{array}{cccc}
3 & 0 & 0 & 0 \\
0 & 1 & 0 & 0\\
0 & 0 & -1 & 0 \\
0 & 0 & 0 & -3
\end{array}\right].
\end{align}
The five Dirac's $\gamma$-matrices are defined in the $4 \times 4$ pseudospin space as
\begin{align}
\gamma^1=& \frac{1}{\sqrt{3}} (J_x J_y +J_y J_x), \;
\gamma^2= \frac{1}{\sqrt{3}} (J_y J_z +J_z J_y), \\
\gamma^3=& \frac{1}{\sqrt{3}} (J_z J_x +J_x J_z), \;
\gamma^4= \frac{1}{\sqrt{3}} (J_x^2- J_y^2), \\
\gamma^5=& \frac{1}{3} (2J_z^2 - J_x^2 -J_y^2), \label{eq:gamma5}
\end{align}
and $1_{4\times 4}$ denotes the identity matrix.
They satisfy the following relations:
\begin{align}
    &\gamma^\mathrm{i}\, \gamma^\mathrm{j} + \gamma^\mathrm{j}\, \gamma^\mathrm{i} =2 \times  1_{4\times 4} \delta_{\mathrm{i}, \mathrm{j}}, \\
    &\gamma^1\, \gamma^2\, \gamma^3\, \gamma^4\, \gamma^5=-1_{4\times 4},\\ 
    &\{\gamma^\mathrm{i}\}^\ast = \{\gamma^\mathrm{i}\}^{\mathrm{T}}= U_T\, \gamma^\mathrm{i} \, U_T^{-1},\quad U_T=\gamma^1\, \gamma^2,
\end{align}
where $U_T$ is the unitary part of the
time-reversal operator $\mathcal{T}=U_T \, \mathcal{K}$, with $\mathcal{K}$ denoting complex conjugation.

\end{document}